\documentclass[pre,twocolumn,10pt,aps,longbibliography]{revtex4-1}
\usepackage{amssymb}
\usepackage{amsmath}
\usepackage{graphicx}% Include figure files
\usepackage{dcolumn}% Align table columns on decimal point
\usepackage{bm}% bold math

\begin{document}

%\preprint{APS/123-QED}

\title{Integrated information in the thermodynamic limit}% Force line breaks with \\
%\title{Manuscript Title:\\with Forced Linebreak}% Force line breaks with \\

\author{Miguel Aguilera}
 \email{sci@maguilera.net}
 \altaffiliation[Also at ]{ISAAC Lab, Arag\'on Institute of Engineering Research, University of Zaragoza, Zaragoza, Spain.}%Lines break automatically or can be forced with \\
\author{Ezequiel Di Paolo}%
 \altaffiliation[Also at ]{Ikerbasque, Basque Foundation for Science, Bizkaia, Spain, and the Centre for Computational Neuroscience and Robotics, Department of Informatics, University of Sussex, Brighton, UK.}
\affiliation{%
IAS-Research Center for Life, Mind, and Society, University of the Basque Country, Donostia, Spain}%

%\author{Ann  Author}
%\affiliation{%
%\affiliation{

\date{\today}% It is always \today, today,

\begin{abstract}
The capacity to integrate information is a prominent feature of biological and cognitive systems. Integrated Information Theory (IIT) provides a mathematical approach to quantify the level of integration in a system, yet its computational cost generally precludes its applications beyond relatively small models. In consequence, it is not yet well understood how integration scales up with the size of a system or with different temporal scales of activity, nor how a system maintains its integration as its interacts with its environment.
Here, we show for the first time how measures of information integration scale when systems become very large. Using kinetic Ising models and mean-field approximations from statistical mechanics, we show that information integration diverges in the thermodynamic limit at certain critical points. 
Moreover, by comparing different divergent tendencies of blocks of a system at these critical points, we delimit the boundary between an integrated unit and its environment.
Finally, we present a model that adaptively maintains its integration despite changes in its environment by generating a critical surface where its integrity is preserved.
We argue that the exploration of integrated information for these limit cases helps in addressing a variety of poorly understood questions about the organization of biological, neural, and cognitive systems.
%An article usually includes an abstract, a concise summary of the work
%covered at length in the main body of the article. It is used for
%secondary publications and for information retrieval purposes. Valid
%PACS numbers may be entered using the \verb+\pacs{#1}+ command.
\end{abstract}

\maketitle

\section{Introduction}

Cognition emerges from the distributed activity of many neural, bodily, and environmental processes.
The problem of large-scale integration of neural processes is crucial for understanding how unified cognitive and behavioural states arise from the coordination of these distributed sources of activity. 
Evidence \cite{bassett_understanding_2011, pessoa_understanding_2014} suggests this integration process is non-decomposable: we cannot understand it in terms of modular components or timescales of activity in a neural system nor can we decouple neural activity from the external environment \cite{aguilera_situated_2013}. The different components and scales of the cognitive process are deeply intertwined. Yet, the functional components of the process are still able to maintain their differentiated characteristics in order to generate complex adaptive patterns of behaviour.

How can such an integrated, complex organization emerge and be maintained? One of the most attractive theories is that neural activity is coordinated into a coherent yet flexible `dynamic core' \cite{varela_resonant_1995, tononi_consciousness_1998}, which balances opposing tendencies of integration and segregation. The interplay of these opposing tendencies generates information (understood as described by information theory, not in a semantic or intensional sense) that is highly diversified among functional parts of the nervous system, and at the same time unified into a coherent whole, thus displaying highly complex patterns of activity.

Integrated information is defined as the information possessed by a system which is above and beyond the information that is available from the sum of its parts. Information integration was first conceived of as linked to consciousness \cite{tononi_consciousness_1998, oizumi_phenomenology_2014} but it can also be manifested without awareness \cite{mudrik_information_2014} and has been used more generally to describe biological autonomy \cite{marshall_how_2017}. Although the topic of information integration has received interest from different communities in recent years, we are still lacking a full understanding of the principles that underlie this fundamental process: how integrative forces are deployed temporally or spatially, how they cope with the surrounding environment, or how they scale with the size of the system.

Different approaches have proposed ways to formalize this idea; one of the most popular has been developed as a measure connected to consciousness under the name of \emph{integrated information theory} (IIT, \cite{oizumi_phenomenology_2014}).
In its latest versions, IIT is based on interventionist notions of causality to characterize the causal influences between the components of a system  \cite{oizumi_phenomenology_2014, marshall_how_2017}. That is, instead of assessing whether a system is unified into a coherent whole by analysing its behaviour in regular conditions, IIT proposes that the forces integrating the behaviour of the system are better captured by observing its behaviour under perturbations.

IIT postulates that any subset of elements of the system is a mechanism~\footnote{We use the term `mechanism' in the technical sense described later and not in the specific sense of efficient causality of the mechanical kind. We acknowledge that different forms of causal and enabling relations between processes are possible and relevant, yet we retain the term `mechanism' in this context to remain coherent with the existing literature.} integrating information if its intrinsic cause-effect power (i.e., its ability to determine past and future states) is irreducible.
Irreducibility is measured in terms of integrated information $\varphi$, which when larger than 0 indicates that the subset of elements at its current state constrains the past and future states of the system in a way that cannot be decomposed in two or more independent cause-effect sets of relations.
That is, $\varphi$ captures the level of irreducibility of the system, understood in the sense that even the least disrupting bipartition of the system into two disconnected halves (this is called the minimum information partition, MIP) would imply a loss of information in the causal power of the system.
Aside from computing integrated information at the level of mechanisms, IIT postulates a composite measure $\Phi$, which is computed from the set of all mechanisms (each one defined by a value of $\varphi$) computed in the original system and the system under bidiriectional partitions. 
A system with $\Phi>0$ is described as forming an irreducible unitary whole. Since many subsets of the system may present $\Phi>0$, the boundaries of the system are defined around the subset with larger $\Phi$. A detailed description of IIT measures is provided in Appendix~\ref{app:IIT}.

Nevertheless, current formulations of IIT present some limitations for studying brain organization. We propose that, in order to extend current uses of IIT to capture some important aspects of neural organization, we should re-examine some of the main assumptions behind its conception:
\begin{itemize}
\item \textbf{Scalability}. A system can present different levels of integration at different spatial and temporal scales \cite{hoel_can_2016, marshall_black-boxing_2018} and, in general, it is not well understood how integration behaves at different scales. However, analyses of the properties of brain-inspired statistical mechanical models have unveiled how many processes in neural systems take the form of phase transitions occurring in the thermodynamic limit, showing properties that diverge as the size of the system scales up. Here we apply models from statistical mechanics to describe integration in terms of the tendencies of the system near the thermodynamic limit. 
\item \textbf{Temporal deployment} The latest formulations of IIT \cite{oizumi_phenomenology_2014} attempt to capture the dynamical nature of neural systems by focusing on the dynamics of causal processes, not taking the stationarity or ergodicity of the system as initial assumptions. Nevertheless, IIT is only measured at a single scale of temporal activity, since it analyses integration in the causal power of a mechanism from one time step to the next. We propose a modification of $\varphi$ to study integration along different temporal spans, showing that systems at critical points must be evaluated for very long timescales.
\item \textbf{Non-decomposability}. As we mentioned, empirical evidence points to the non-decomposability of cognitive processes. In its current formulation, IIT considers elements outside the system under analysis as independent sources of noise. Here, we propose instead that the level of integration of a system must be evaluated in the context of the other systems it is coupled to (therefore not assuming that elements in the environment are just sources of statistical noise). This modification allows us to correctly determine the boundary between a system and its environment in the thermodynamic limit.
\end{itemize}
Some of the assumptions and modifications pointed out here are explained later in the text, and a detailed account and comparison between IIT and our measure of integrated information can be found in Appendix~\ref{app:phi}.
Part of the reasons why some of the aspects above have not yet been addressed is that, due to its computational complexity, the application of current IIT measures is limited to very small systems and short timescales. In general, IIT has been tested in small toy models (e.g., \cite{oizumi_phenomenology_2014,albantakis_evolution_2014}, although some alternative formulations try to circumvent this problem, see \cite{barrett_practical_2011,oizumi_measuring_2016}). 
In contrast, our approach, apart from the modifications proposed above, introduces some simplifications and approximations in order to measure integrated information as a system scales to very large sizes. Specifically, we introduce a simple kinetic Ising model of infinite size and quasi-homogeneous connectivity, which presents an exact mean field solution that we use to simplify the calculation of integrated information $\varphi$ of the mechanisms of a system.

We proceed as follows. First, we introduce the kinetic Ising model and a mean field approximation for solving it. 
Then, we introduce a measure of integrated information and how it can be computed for Ising models of infinite size.
Finally, we present the results of our method in three scenarios of increasing complexity for depicting how integrated information can be used to characterize an integrated system interacting with an environment:
\begin{itemize}
\item In the first scenario, we illustrate the measure in a simple homogeneous model. In the thermodynamic limit, we can describe integrated information as the susceptibility of the system to changes in the direction of the minimum information partition (MIP). Consequently, integrated information diverges when the system is near a critical point.
\item The second scenario depicts a system coupled to an external environment, showing the system and the system-environment compound both show integrated information diverging near a shared critical point. Nevertheless, depending on the coupling strength, the system and system-environment mechanisms present different speeds of divergence. This allows us to delimit the dominant dynamical unit where integration takes place.
\item Finally, we tune the parameters of a system with internal self-regulation in order to present high integration when interacting with a variety of environments. The system's internal inhibitory interactions generate a critical surface in the direction of the MIP which describe the viable region in which its integration is maintained.
\end{itemize}
The results presented here represent a first attempt at using integrated information theory to delimit the boundaries of a family of infinite size systems that can be formally solved. The interest of the study is twofold. First, it allows us to check some of the assumptions of IIT and propose some modifications to maintain its consistency in the thermodynamic limit, and to propose a way to adapt IIT measures for very large systems. 
Second, although the results presented are obtained from relatively simple cases, they offer an opportunity to speculate about how the causal integrative forces of a system (both its internal cohesion and the coupling with its environment) might scale up when a system approaches the thermodynamic limit. This provides an opportunity to address unanswered questions about integrated organization of biological and cognitive systems.

\section{Model}

We start by describing a general model defining causal temporal interactions between variables.
Looking for generality, we use the least structured statistical model (i.e., a maximum caliber model \cite{presse_principles_2013}) defining causal correlations between pairs of units from one time step to the next.
We study a kinetic Ising model where $N$ binary variables (Ising spins) $s_i$ evolve in discrete time, with synchronous parallel dynamics (Fig~\ref{fig:ising}.A). Given the configuration of spins at the previous step, $s(t-1)=\{s_1(t-1),\dots,s_N(t-1)\}$, the spins $s_i(t)$ are independent random variables drawn from the distribution:
\begin{equation}
  p(s_i(t)|s(t-1))= \frac{e^{\beta s_i(t) h_i(t)}}{2 \cosh(\beta h_i(t))}
%  p(s_i(t)|s(t-1))= \frac{1}{1 + e^{-2\beta s_i(t) h_i(t)} }
%  p(s_i(t)|s(t-1))= [1 + e^{-2\beta s_i(t) h_i(t)} ]^{-1}
 \label{eq:Ising}
\end{equation}
where
\begin{equation}
h_i(t)=H_i+\sum_j J_{ij} s_j(t-1)
 \label{eq:Ising-field}
\end{equation}
The parameters $H_i$ and $J_{ij}$ represent the local fields at each spin and the couplings between pairs of spins, and $\beta$ is the inverse temperature of the model.
Without loss of generality, we assume $\beta=1$.

\subsection{Mean field kinetic Ising model}
We focus on the particular case of a system of infinite size where $H_i=0$. The system is divided into different regions (from 1 to 3 depending on the example), and the coupling values $J_{ij}$ are positive and homogeneous for each intra- or inter-region connections $J_{ij} =\frac{1}{N_{\mathcal{R}}} J_{\mathcal{S} \mathcal{R}}$, where $\mathcal{R}$ and $\mathcal{S}$ are regions of the system with sizes $N_{\mathcal{R}},N_{\mathcal{S}}$ and $i\in\mathcal{S}, j\in\mathcal{R}$. 

For a system of infinite size (and all regions with also infinite size), a mean field approximation allows to calculate the field of all units $i$ belonging to the region $\mathcal{S}$ as: 
\begin{equation}
\begin{aligned} 
 h_i(t)=\sum_\mathcal{R} J_{\mathcal{S} \mathcal{R}} m_\mathcal{R}(t-1), \\
 m_\mathcal{R}(t-1) = \frac{1}{N_{\mathcal{R}}} \sum_{j\in\mathcal{R}} s_j(t-1)
 \end{aligned}
 \label{eq:mean-field}
\end{equation}
where $m_\mathcal{R}(t-1)$ is the mean field of region $\mathcal{R}(t-1)$.
Now we can exactly define the update of the mean field variables using Eq~\ref{eq:Ising} as:
\begin{equation}
m_\mathcal{S}(t)=\tanh(\sum_\mathcal{R} J_{\mathcal{S} \mathcal{R}} m_\mathcal{R}(t-1))
\label{eq:update-mean-field}
\end{equation}

\subsection{Integrated Information $\varphi$}

We use a simplified version of the integrated effect information described by IIT \cite{oizumi_phenomenology_2014}, implementing some modifications to measure the scaling of integrated information in the thermodynamic limit. 
In IIT, both causes and effects of a state are taken into account. For simplicity, we consider only the effects of a particular state. Also, although IIT is defined only for the immediate effects after one update of the state of the system, we define integrated information $\varphi(\tau)$ for an arbitrary number of updates of the system. See Appendix~\ref{app:phi} for a list of the differences between IIT and the measure employed here.

\begin{figure}
 \includegraphics[width=8.0cm]{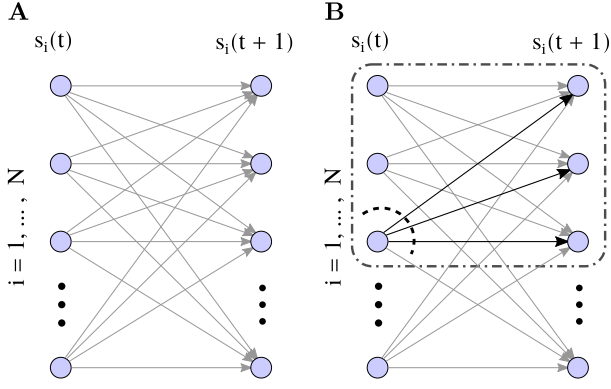}
 \caption{{\bf Kinetic Ising model.}
 A: Description of the infinite size kinetic Ising model. B: Description of the partition schema used to define perturbations. Partitioned connections (black arrows) are injected with random noise.}
 \label{fig:ising}
 \end{figure}

Given an initial state $s(\tau_0)$, we define a `mechanism' $\mathcal{M}$ (following IIT's nomenclature) as a subset of units $\{s_i(\tau_0)\}_{i \in \mathcal{M}}$.
The integrated information of mechanism $\mathcal{M}$, $\varphi_\mathcal{M}$, is defined as the distance between the behaviour of the original system to a system in which a partition (from the set of possible bipartitions) is applied over the units in $\mathcal{M}$.
Fig~\ref{fig:ising}.B depicts an example of a partition.
When a partition is applied, the input coming from the partitioned connections of the system is replaced by a random unconstrained noise (binary white noise in the case of an Ising model).

Once the partition is applied, the probability of the state $s(\tau_0+\tau)$ is computed after $\tau$ updates, injecting noise at the partitioned elements during each update. Then, integrated information is defined as the distance $D$ between the conditional probability distributions at $t+\tau$:
\begin{equation}
\varphi_\mathcal{M}^{cut}(\tau)=D(p(s(\tau_0+\tau)|s(\tau_0)) , p^{cut}(s(\tau_0+\tau)|s(\tau_0)))
\end{equation}
where $D(p_1,p_2)$ refers to the Wasserstein distance (also known as earth mover's distance) used by IIT to quantify the statistical distance between probability distributions.
Here $cut$ specifies the partition applied over the elements of mechanism $\mathcal{M}$, $cut=\{ S_1^c,S_2^c,S_1^f,S_2^f\}$, where $S_1^c,S_2^c$ design the blocks of a bipartition of the mechanism at the current state $\{s_i(t)\}_{i\in \mathcal{M}}$, and $S_1^f,S_2^f$ refer to the blocks of a bipartition (not necessarily the same) of the updated state of the units $\{s_i(t+1)\}_{i\in\mathcal{M}}$. 
Fig~\ref{fig:ising}.B  represents the partition $cut=\{\{ s_1(t),s_2(t)  \},\{s_3(t) \},\{ s_1(t+1),s_2(t+1),s_3(t+1) \},\{  \}  \}$.

Specifically, IIT computes integrated information as the value of $\varphi^{cut}$ under the minimum information partition (MIP), which is the partition of mechanism with the least difference to the original partition (i.e., $\varphi^{MIP}_M(\tau)= \min_{cut} \varphi^{cut}_M(\tau)$).
We use $\varphi_\mathcal{M}(\tau)$ to denote the minimum information partition integrated information $\varphi^{MIP}_M(\tau)$.

Note that some important modifications have been made. 
The most important one is that IIT considers the element outside of the mechanism as unconstrained sources of noise. As we show in Figure~\ref{fig:iit-external-activity}, this can radically change the results of integrated information theory, provoking spurious divergences at points other than the critical point. To preserve the consistency of our results, we let elements outside the mechanism operate normally (see Appendix~\ref{app:phi-environment} for details).

%\begin{figure}[ht]
%\begin{center}
% \begin{tabular}{cc}
%   \multicolumn{1}{l}{\textbf{A}} & \multicolumn{1}{l}{\textbf{B}}\\
%  \includegraphics[width=4.0cm]{img/fig-ising.eps} & \includegraphics[width=4.0cm]{img/fig-partition.eps}%\\
% \end{tabular}
%\end{center}
%\caption{\textbf{(A)} Description of the infinite size kinetic Ising model. \textbf{(B)} Description of the partition schema used to define perturbations. Partitioned connections (black arrows) are injected with random noise.} 
%\label{fig:ising}
%\end{figure}

\subsection{Integrated information in the mean field model}

We now show how integrated information can be computed for the mean field approximation of the Ising model.
Thanks to the mean field approximation we can simplify the calculation of the probability distributions of trajectories $p(s(\tau_0+\tau)|s(\tau_0)) , p^{cut}(s(\tau_0+\tau)|s(\tau_0))$ to a Markovian distribution dependent on the mean field at the previous step.

In general, $p(s(\tau_0+\tau)|s(\tau_0))$ can be computed recursively applying the equation:
\begin{equation}
\begin{aligned}
p(s(\tau_0+\tau)|p(\tau_0))  = \\ \sum\limits_{s(\tau_0+\tau -1)} p(s(\tau_0+\tau )|s(\tau_0+\tau -1)) p(s(\tau_0+\tau -1)|s(\tau_0))
\end{aligned}
\end{equation}
In the kinetic Ising model of inifine size, the mean fields of the system's regions are deterministic, and instead of computing all possible paths of the system we can just determine the evolution of the mean field using Equation~\ref{eq:update-mean-field}. 
Moreover, knowing the mean field of each region we can calculate the value of the effective fields  $h(\tau_0+\tau)$ received by each unit using Equation~\ref{eq:mean-field}.
Also, given the mean field value at a specific point, the posterior probability distribution of each unit is independent. Thus, using the value of $h(\tau_0+\tau)$ computed evolving from $s(\tau_0)$ we can just take: 
\begin{equation}
\begin{aligned}
p(s_i(\tau_0+\tau)|s(\tau_0)) = p(s_i(\tau_0+\tau)|h_i(\tau_0+\tau))
\end{aligned}
\end{equation}

In this context, the calculation of the Wasserstein distance $D$ is drastically simplified, and we can compute $\varphi$ as the sum of distances between independent binary variables, which is equivalent to computing the difference of their mean values:
\begin{equation}
\begin{aligned}
\varphi_\mathcal{M}^{cut}(\tau) = 
%\sum_i D( p(s_i(\tau_0+\tau)|h_i(\tau_0+\tau)), p^{cut}(s_i(\tau_0+\tau)|h_i(\tau_0+\tau))) \\ =
\frac{1}{2} \sum_\mathcal{R} N_\mathcal{R}\lvert m_\mathcal{R}(\tau_0+\tau) - m_\mathcal{R}^{cut}(\tau_0+\tau) \rvert 
\end{aligned}
\end{equation}

Once we can calculate $\varphi$, we still have the problem of finding the MIP of the system.
Luckily, since the connectivity of the system is homogeneous for all nodes in the same region, finding the MIP is equivalent to finding the partition that cuts the lowest number of connections. For infinite size systems where inter-region connections are not zero, the MIP will be one of the possible partitions that isolate just one node of the system.
Also, the partition that isolates a single unit in time $t$ always has a smallest value of $\varphi$ than the partition isolating a node at time $t+1$, since partitioning the posterior distribution corresponds to a larger difference between $m_\mathcal{R}(\tau_0+\tau)$ and $m_\mathcal{R}^{cut}(\tau_0+\tau)$.
Thus, finding the MIP corresponds to finding which region $\mathcal{R}$ of the system least affects future states when one node of the region is isolated in the partition at time $t$ (e.g., Fig~\ref{fig:ising}.B).

Finally, we define a function $F_\mathcal{R}(m(\tau_0),\tau,\{J_{\mathcal{S},\mathcal{R}}\})$ that recursively applies the update rule in Eq~\ref{eq:update-mean-field} for $\tau$ steps starting from an initial value with a mean field value $m(\tau_0)$, such that $m_{\mathcal{R}}(\tau_0+\tau) = F_{\mathcal{R}}(m(\tau_0),\tau,J)$. In our mean field approximation, applying the MIP to the quasi-homogeneous system described here is equivalent to just removing one connection~\footnote{Note that cutting a connection implies injecting uniform noise, which in the mean field approximation is equivalent to substitute the input by a zero mean field or just removing the connection. This is an important approximation that allow us to obtain the main results of the paper, although it will only be valid when the size of the system is infinite and $\tau$ is larger than $1$.} between one or more pairs of regions $\{\mathcal{S},\mathcal{R}\}_{cut}$, whereas the connections between the rest of regions $\{\mathcal{S},\mathcal{R}\}_{uncut}$ remain intact.
Therefore the update rule applied by function $F$ to the partitioned system is $F(m(\tau_0),\tau,\{\{J_{\mathcal{S},\mathcal{R}}\}_{uncut},\{(1-\frac{1}{N_\mathcal{R}}) J_{\mathcal{S},\mathcal{R}}\}_{cut}\})$.

Assuming that the number of units per region is equal to $N_\mathcal{R} = r_\mathcal{R} N$ and  $\sum {r_\mathcal{R}} = 1$, we get a simplified expression for the partitioned and unpartitioned terms:
\begin{equation}
\begin{aligned}
F_{\substack{\mathcal{R}\\ cut}}(m_0,\tau,x) \\ =
 F_\mathcal{R}(m_0,\tau,\{\{J_{\mathcal{S},\mathcal{R}}\}_{uncut},\{(1-\frac{x}{r_\mathcal{R}}) J_{\mathcal{S},\mathcal{R}}\}_{cut}\}) \rvert
\end{aligned}
\end{equation}
where $m_0=m(\tau_0)$ and $x=\frac{1}{N}$ in the partitioned case and $x=0$ otherwise. Now, computing the unpartitioned and partitioned cases case is equivalent to calculating $F_{\substack{\mathcal{R}\\ cut}}(m_0,\tau,0)$ and $F_{\substack{\mathcal{R}\\ cut}}(m_0,\tau,\frac{1}{N})$ respectively.
Given this, assuming $N\to\infty$ we calculate the final form of $\varphi$ as a sum of the derivatives of function $F_{\substack{\mathcal{R}\\ cut}}(m_0,\tau,x)$:
\begin{equation}
\begin{aligned}
\varphi_\mathcal{M}^{cut}(\tau)  \\ =
\lim_{N\to\infty} \frac{1}{2} \sum_\mathcal{R} N_{\mathcal{R}} \lvert F_{\substack{\mathcal{R}\\ cut}}(m_0,\tau,0) - F_{\substack{\mathcal{R}\\ cut}}(m_0,\tau,\frac{1}{N})\rvert \\ =
% \frac{1}{2} \sum_{\mathcal{R}} \lvert - N_{\mathcal{R}}  F'_{\substack{\mathcal{R}\\ cut}}(m_0,\tau,0) \frac{1}{N} \rvert = \\
\frac{1}{2} \sum_{\mathcal{R}} \lvert r_{\mathcal{R}}  F'_{\substack{\mathcal{R}\\ cut}}(m_0,\tau,0) \rvert 
\end{aligned}
\end{equation}
where $ F'(m_0,\tau,x) = \frac{d F'(m_0,\tau,x) }{dx}$. Note that this defines integrated information in similar terms as the magnetic susceptibility typically used in Ising model to identify critical points, although in this case the mean field of the system is differentiated along the parametrical direction of the MIP.

\section{Results}

\subsection{Integrated information in a homogeneous kinetic Ising model}

As an example, we compute numerically the value of $\varphi_{\mathcal{M}_N}(\tau)$ for a homogeneous kinetic Ising model containing just one region (as in Fig~\ref{fig:ising}.A). The system only has one parameter $J$ describing all connections in the system.

\begin{figure}
	\includegraphics[width=8.5cm]{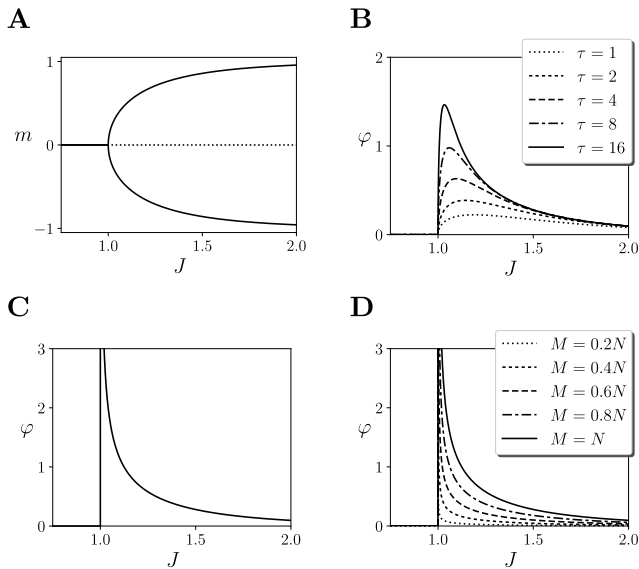}
\caption{{\bf Homogeneous kinetic Ising model.}
A: Magnetization of the infinite size kinetic Ising model. B: Value of $\varphi_{\mathcal{M}_N}(\tau)$ for different temporal spans. C: Value of $\varphi_{\mathcal{M}_N}(\tau\to\infty)$ for an infinite temporal span. D: Value of $\varphi_{\mathcal{M}_M}(\tau\to\infty)$ for different mechanisms of size $M$ and an infinite temporal span.}
\label{fig:phi-infinite}
\end{figure}

For different values of $J$, we compute $\varphi$ for the system starting from a state in the stationary solution. For doing so, we need to know how to compute $F_{cut}(m_0,\tau,x)$, that is, how to compute the mean field of units at a particular time.

First, we numerically compute  $F_{cut}(m_0,\tau,x)$ and $\varphi_{\mathcal{M}_N}$ for different values of $J$ for the largest mechanism $\mathcal{M}_N$ of size $N$, and different values of $\tau$ and $m(\tau_0)$ equal to the value at the stationary solution of the system. We estimate the values of the derivative as $F'_{cut}(m_0,\tau,0) = (F_{cut}(m_0,\tau,dx)-F_{cut}(m_0,\tau,0))/dx$, using a value $dx=10^{-10}$. 
As we observe in Fig~\ref{fig:phi-infinite}.B, the value of $\varphi_{\mathcal{M}_N}(\tau)$ appears to diverge as $\tau$ grows~\footnote{ Note that for larger $\tau$ the partition is applied for a longer period of time, and therefore yielding larger integration in some cases.}.

Similarly, we numerically compute $\varphi_{\mathcal{M}_N}(\tau\to\infty)$  by using the mean field of the model iterating the equation $m(t)=tanh(J m(t-1))$ until the difference in the update is smaller than $10^{-15}$. In Fig~\ref{fig:phi-infinite}.C we observe how $\varphi_{\mathcal{M}_N}(\tau\to\infty)$  shows an apparent divergence around $J=1$. 
Also, we compute the value of $\varphi_{\mathcal{M}_M}(\tau\to\infty)$ for different mechanisms of size $M$ as a fraction of $N$. As shown in Fig~\ref{fig:phi-infinite}.D, the resulting value of integrated information still diverges but is smaller than the value of $\varphi_{\mathcal{M}_N}(\tau)$ of the whole system, indicating that the system is irreducible.

We can go beyond numerical computations and calculate the analytic value of  $\varphi_{\mathcal{M}_N}(\tau\to\infty)$ near the point of divergence by approximating the values of $F_{cut}(m_0,\tau\to\infty,0)$ around $J=1$ as the value of $m$ that solves \mbox{$m=tanh(Jm)$}.
Note that, more generally, we can compute $F_{cut}(m_0,\tau\to\infty,x)$ just by substituting $J\leftarrow J(1-x)$.

The system  has a trivial solution at $m=0$. Also, for $J>1$ the solution at $m=0$ becomes unstable and a pair of solutions in a pitchfork bifurcation (Fig~\ref{fig:phi-infinite}.A).
Although there is no analytic solution of the problem, we can compute the value of $m$ near $J=1$ by approximating the hyperbolic tangent by the first two terms of its Taylor series, finding that in the limit $J\to1^+$ we approximate: 
\begin{equation}
\begin{aligned}
 F_{cut}(m_0,\tau\to\infty,x)=\pm \sqrt{\frac{3(J(1-x)-1)}{(J(1-x))^3}}\\
 \varphi_{\mathcal{M}_N}(\tau\to\infty) = 
%  \lvert F'_{cut}(m_0,\tau\to\infty,0) \rvert =
  \frac{1}{2} \bigg\lvert \dfrac{\sqrt{3}\left(2J-3\right)}{2\sqrt{J^3(J-1)}} \bigg\rvert
 \end{aligned}
\end{equation}
Thus, we can confirm that the value of integrated information $\varphi_{\mathcal{M}_N}(\tau\to\infty) $ diverges when $J\to1^+$. 
This has interesting implications.
If the a system must maintain a growing level of integration as its size increases, it needs to be poised near a critical point that shows a divergence of the values of $\varphi$.

%\begin{figure}
%\begin{center}
% \begin{tabular}{cc}
%   \begin{tabular}{c}
%     \multicolumn{1}{l}{\textbf{A}} \\
%     \includegraphics[width=37mm]{img/mf-infinite-size.eps}
%   \end{tabular}
%    &
%   \begin{tabular}{c}
%     \multicolumn{1}{l}{\textbf{B}} \\
%     \includegraphics[width=37mm]{img/phi-infinite-size-finite-time.eps}
%   \end{tabular}
%   \\
%   \begin{tabular}{c}
%     \multicolumn{1}{l}{\textbf{C}} \\
%     \includegraphics[width=37mm]{img/phi-infinite-size.eps} 
%   \end{tabular}
%    &
%   \begin{tabular}{c}
%     \multicolumn{1}{l}{\textbf{D}} \\
%     \includegraphics[width=37mm]{img/phi-infinite-size-mechanisms.eps}
%   \end{tabular}
% \end{tabular}
%\end{center}
%\caption{\textbf{(A)} Magnetization of the infinite size kinetic Ising model. \textbf{(B)} Value of $\varphi_{\mathcal{M}_N}(\tau)$ for different temporal spans. \textbf{(C)} Value of $\varphi_{\mathcal{M}_N}(\tau\to\infty)$ for an infinite temporal span. \textbf{(D)} Value of $\varphi_{\mathcal{M}_M}(\tau\to\infty)$ for different mechanisms of size $M$ and an infinite temporal span.} 
%\label{fig:phi-infinite}
%\end{figure}

\subsection{Integrated information for measuring agent-environment asymmetries}

We apply the proposed measure of integrated information to the problem of determining the boundaries of an agent interacting with an environment.
One of the central aspects of agency is the existence of agent-environment asymmetries \cite{barandiaran_defining_2009}, in which the part of the system corresponding to the agent is able (to an extent) to define the terms in which it relates to the surrounding milieu.
We test our measure in two simple cases of systems presenting asymmetries in their interaction.

\begin{figure*}
	\includegraphics[width=16.0cm]{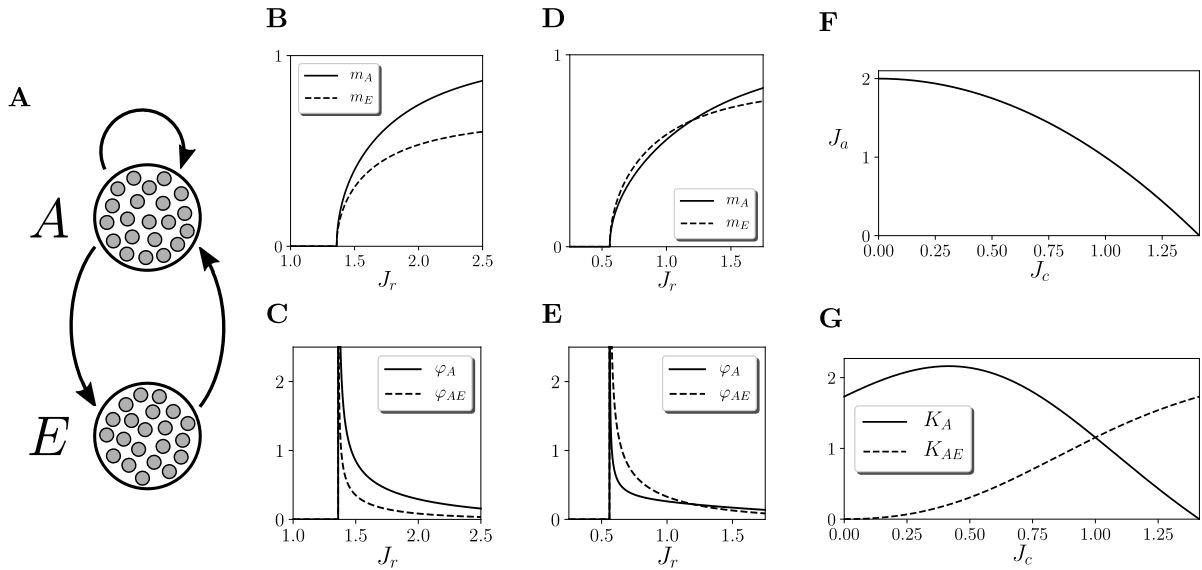}
\caption{{\bf Asymmetric interaction in a kinetic Ising model.}
A: Basic agent connected to an environment.  B, C, D, E: Values of the mean fields (only positive values are shown) of the stable solution (top) and  $\varphi(\tau\to\infty)$  (bottom) for the agent and environment nodes of the model at stability for $J_c=0.8$ (left) and $J_c=1.2$ (right) and different values of $J_r$. F: location of the critical point in the parameter space for different combinations of $J_r,J_c$. 
G: Constants multiplying $\varphi_A(\tau\to\infty)$ and $\varphi_{AE}(\tau\to\infty)$ near the critical point, showing which is the most irreducible unit of the system.}
\label{fig:agent-environment}
\end{figure*}

We model a minimal case of agent-environment bidirectional interaction with two regions, where only the region corresponding to the `agent' has the capacity to self-regulate through recurrent connections (Fig~\ref{fig:agent-environment}.A).
In this case, we have two regions $A$ and $E$, only $A$ presenting self-connections. The mean field of the system is updated as:
\begin{equation}
 \begin{aligned}
 m_A(t+1) = \tanh(\frac{1}{2}( J_{AA} m_A(t) + J_{AE} m_E(t))) \\
 m_E(t+1) = \tanh(J_{EA} m_A(t))
 \end{aligned}
 \label{eq:agent-environment}
\end{equation}
For simplicity, we study the case where agent-environment connections are symmetric $ J_{AE}= J_{EA}=J_c$, and $J_{AA}=J_r$.
We numerically compute that the system has an similar solution than the previous case, presenting a pitchfork bifurcation at a critical point (Fig~\ref{fig:agent-environment}.B,D).

Moreover, we compute the value of $\varphi_\mathcal{M}(\tau\to\infty)$ for different mechanisms.
For the case of the mechanism covering the whole system $\mathcal{M}=AE$, we look for the MIP of the system by isolating single units of the mechanism at $s(t)$ (Fig~\ref{fig:ising}.B). 
If we isolate a unit from region $A$, two connections are cut (one with value $J_r$ and one with value $J_c$). Otherwise, if we isolate a unit from region $E$, only one connection with value $J_c$ is cut. Thus, this second partition is  always the MIP of the system ($MIP_{AE}$).
For $\mathcal{M}=A$, the only candidate for the MIP is isolating one node from $A$, therefore cutting one connection with value $J_r$ ($MIP_{A}$). Finally, for mechanism $E$ there are no connections within the mechanism and we can directly conclude that $\varphi_E=0$.

Now, the question is: can we consider $A$ as an individual system or should we consider instead the coupled system $AE$ as an integrated unit?
Assuming $r_A=r_E=0.5$, we define the values of integrated information as:
\begin{equation}
 \begin{aligned}
 \varphi_A = \frac{1}{4} ( \lvert  \sum_{\mathcal{R}={A,E}} F'_{\substack{\mathcal{R}\\ MIP_{A}}}(m_0,\tau,0) \rvert )\\
 \varphi_{AE} = \frac{1}{4} ( \lvert  \sum_{\mathcal{R}={A,E}} F'_{\substack{\mathcal{R}\\ MIP_{AE}}}(m_0,\tau,0) \rvert )
 \end{aligned}
\end{equation}

In Fig~\ref{fig:agent-environment}.C,E we estimate the value of $\varphi_A, \varphi_{AE}$ for $\tau\to\infty$ an initial value $m_0$ corresponding to the stationary solution of the system, and values of $J_c=0.8$ (left) and $J_c=1.2$ (right).
We observe that in all cases the values of $\varphi_A,\varphi_{AE}$ diverge next to the critical point. 
Nevertheless, in the first case when agent-environment connections are weaker $\varphi_A>\varphi_{AE}$ next to the critical point. In contrast, for stronger couplings between agent and environment $\varphi_A<\varphi_{AE}$ in the vicinity of the critical point.

We validate this results by solving Eq~\ref{eq:agent-environment} near criticality. We do this by transforming it into a system of one equation $m_A=\tanh(\frac{1}{2}( J_{AA} m_A + J_{AE} \tanh(J_{EA} m_A)))$ and finding its Taylor series near $m_A=0$. 
We obtain that near the critical point:
\begin{equation}
 \begin{aligned}
 F_A(m_0,\tau\to\infty,0)  = \sqrt{\dfrac{3(J_{AA} + J_{AE} J_{EA} -2)}{J_{AE} J_{EA}^3 + \frac{1}{4} (J_{AA}+J_{AE}J_{EA})^3}} \\
 F_E(m_0,\tau\to\infty,0)  = \tanh(J_{EA} F_A(m_0,\tau\to\infty,0) )
 \end{aligned}
\end{equation}
Similarly, \mbox{$F_A(m_0,\tau\to\infty,x)$} and \mbox{$F_E(m_0,\tau\to\infty,x)$} are easily calculated by adding a $(1-x)$ factor to the partitioned connections.
Thus, we find that the location of the critical point which is the one satisfying $J_{AA} + J_{AE} J_{EA} = 2$ (Fig~\ref{fig:agent-environment}.F).
From here, we get:
\begin{equation*}
 \begin{aligned}
   F'_{A|MIP_A} = 
   \frac{3}{2}  \frac{-J_{AA}}{J_{AE} J_{EA}^3 + \frac{1}{4} (J_{AA}+J_{AE}J_{EA})^3}  \\
   \cdot ( \frac{1}{4}(J_{AA}+J_{AE}J_{EA})^2   F_A + \frac{1}{F_A})\\
   F'_{E|MIP_A}   = 
   \frac{J_{EA}}{\cosh(J_{EA} F_A )^2} F'_{A|MIP_A}
  \end{aligned}
\end{equation*}  
\begin{equation*}
 \begin{aligned}
   F'_{A|MIP_{AE}} = 
    \frac{3}{2}  \frac{-J_{AE}J_{EA}}{J_{AE} J_{EA}^3 + \frac{1}{4} (J_{AA}+J_{AE}J_{EA})^3} \\
    \cdot (\frac{J_{EA}^2}{3}  +  \frac{1}{4}(J_{AA}+J_{AE}J_{EA})^2  F_A + \frac{1}{F_A}) \\
   F'_{E|MIP_{AE}}  =
   \frac{J_{EA}}{\cosh(J_{EA} F_A(m_0,\tau\to\infty,0) )^2} F'_{A|MIP_{AE}} 
 \end{aligned}
\end{equation*}
where $F_\mathcal{R}=F_\mathcal{R}(m_0,\tau\to\infty,0)$ and $F'_{\mathcal{R}|MIP_\mathcal{S}}=F'_{\substack{\mathcal{R}\\ MIP_{\mathcal{S}}}}(m_0,\tau\to\infty,0)$.

%\begin{equation}
% \begin{aligned}
%   F'_{\substack{A\\ MIP_{A}}}(m_0,\tau\to\infty,0) = 
%   \frac{3}{2} J_{AA} \frac{1}{J_{AE} J_{EA}^3 + \frac{1}{4} (J_{AA}+J_{AE}J_{EA})^3}  
%   \cdot  ( \frac{1}{4}(J_{AA}+J_{AE}J_{EA})^2  \cdot F_A(m_0,\tau\to\infty,0) - \frac{1}{F_A(m_0,\tau\to\infty,0)})\\
%   F'_{\substack{E\\ MIP_{A}}}(m_0,\tau\to\infty,0)  = 
%   \frac{J_{EA}}{\cosh(J_{EA} F_A(m_0,\tau\to\infty,0) )^2} F'_{\substack{A\\ MIP_{A}}}(m_0,\tau,0)
%  \end{aligned}
%\end{equation}  
%\begin{equation}
% \begin{aligned}
%   F'_{\substack{A\\ MIP_{AE}}}(m_0,\tau\to\infty,0) = 
%    \frac{3}{2}  J_{AE}J_{EE}\frac{1}{J_{AE} J_{EA}^3 + \frac{1}{4} (J_{AA}+J_{AE}J_{EA})^3} 
%    \cdot (\frac{J_{EA}^2}{3}  +  \frac{1}{4}(J_{AA}+J_{AE}J_{EA})^2 \cdot F_A(m_0,\tau\to\infty,0) - \frac{1}{F_A(m_0,\tau\to\infty,0)}) \\
%   F'_{\substack{E\\ MIP_{AE}}}(m_0,\tau\to\infty,0)  =
%   \frac{J_{EA}}{\cosh(J_{EA} F_A(m_0,\tau\to\infty,0) )^2}  \cdot (F'_{\substack{A\\ MIP_{AE}}}(m_0,\tau\to\infty,0) - F_A(m_0,\tau\to\infty,0) )
% \end{aligned}
%\end{equation}
%\normalsize
Near the critical point at $(J_{AA}+J_{AE}J_{EA})\to2^{+}$, the values of integrated information are approximated by the expressions:
\begin{equation}
 \begin{aligned}
  \varphi_A =  J_{AA}  K  (J_{AA}+J_{AE}J_{EA}-2)^{-1/2},\\
  \varphi_{AE} = J_{AE} J_{EA} K (J_{AA}+J_{AE}J_{EA}-2)^{-1/2},\\
  K = \frac{\sqrt{3}(1+J_{EA})}{\sqrt{J_{AE} J_{EA}^3 + \frac{1}{4} (J_{AA}+J_{AE}J_{EA})^3}}
  \end{aligned}
\end{equation}
by defining $K_A=J_{AA} K$ and $K_{AE} = J_{AE} J_{EA} K $ we describe with these variables the level of integrated information of the agent and the whole agent-environment system near the critical point.
In Fig~\ref{fig:agent-environment}.G we observe that there is a transition from the agent being the system with highest integration to the agent-environment. 

This illustrates that, near a critical point, the value of integrated information scales up indefinitely in an agent-environment system. In the case of symmetric interaction only for some cases the agent can be identified as the predominant integrated unit in the system, while in others the agent-environment system is the predominant unit.

%\begin{figure*}
%\begin{center}
% \begin{tabular}{cc}
%   \begin{tabular}{c} 
%   \multicolumn{1}{l}{\textbf{A}} \\
%   \includegraphics[width=4.0cm]{img/fig4a.eps}  \\
%   \multicolumn{1}{l}{\textbf{B}} \\
%   \includegraphics[width=4.5cm]{img/fig4b.eps}  \\
% \end{tabular}
% &
% \begin{tabular}{c} 
%   \multicolumn{1}{l}{\textbf{C}}  \\
%   \includegraphics[width=6.0cm]{img/fig4c.eps}  \\
%   \multicolumn{1}{l}{\textbf{D}}  \\
%   \includegraphics[width=4.0cm]{img/fig4d.eps}
%  \end{tabular}
% \end{tabular}
%\end{center}
%\caption{\textbf{(A)} Adaptive sensorimotor system connected to an environment.  \textbf{(B)}  Values of the mean fields of the stable solution for a $J_c=1$.  \textbf{(C)} Values of  $\varphi_{AB}(\tau)$  for different values of $J_c$. \textbf{(F)} The blue area represents the surface in $J_c$ and $J_{EE}$ where  $\varphi(\tau\to\infty)$ diverges.} 
%\label{fig:agent-sensorimotor}
%\end{figure*}

\subsection{Adaptive integrated information facing environmental diversity}

We have just used integrated information for delimiting an agent interacting with a static environment. The environment was `passive' in the sense that it showed no self-interaction. 
This is not a common scenario, since typically environments change and display their own dynamics.
A key aspect of agency is the ability of an agent to sometimes \emph{modulate} the coupling with its environment to preserve its individuality \cite{barandiaran_defining_2009}, generating an \emph{interactional asymmetry} between agent and environment. Thus, a basic feature of living and cognitive sysetms is to display adaptive mechanisms regulating its coupling to the environment to maintain their level of functional integration for a range of external environments.

\begin{figure*}
\includegraphics[width=16.0cm]{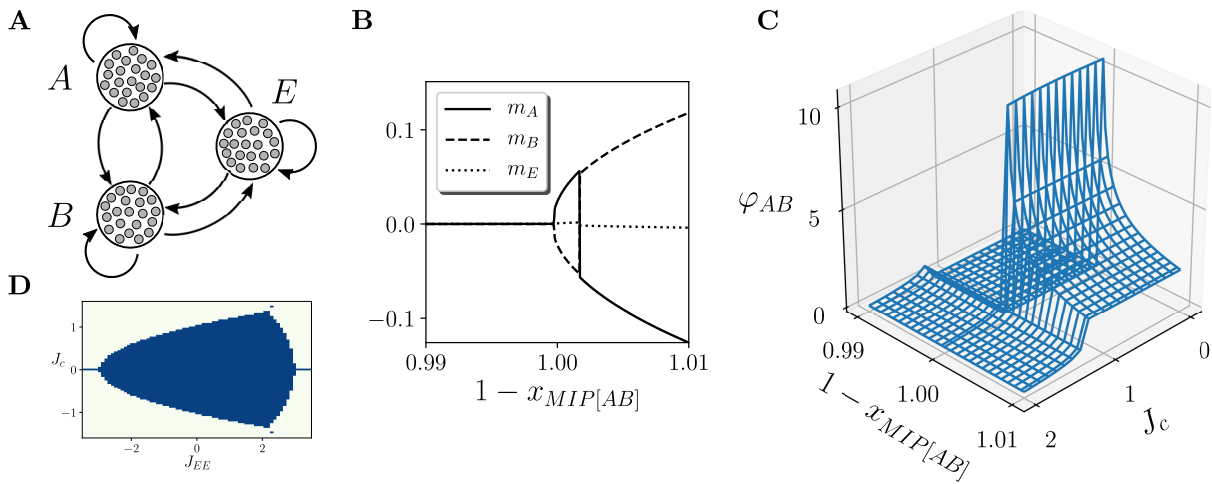}
\caption{{\bf Adaptive integration in a kinetic Ising model.}
A: Adaptive sensorimotor system connected to an environment.  B: Values of the mean fields of the stable solution for a $J_c=1$.  C: Values of  $\varphi_{AB}(\tau)$  for different values of $J_c$. F: The blue area represents the surface in $J_c$ and $J_{EE}$ where  $\varphi(\tau\to\infty)$ diverges.}
\label{fig:agent-sensorimotor}
\end{figure*}

In order to characterize a scenario that is more realistic in this sense, we model an agent with two internal regions $A$ and $B$, interacting with an environment $E$ with recurrent connections (Fig~\ref{fig:agent-sensorimotor}.A). $A$ and $B$ present feedback loops that we fit in order to maintain integration for a range of environmental parametric configurations. The evolution of the system is described by:
% \begin{equation}
% \begin{aligned}
% m_{A}(t+1) =  \tanh(\frac{1}{3}( J_{{A}{A}} m_{A}(t) + J_{{A}{B}} m_{B}(t) + J_{{A}E} m_E(t))) \\
% m_{B}(t+1) = \tanh(\frac{1}{3}( J_{{B}{A}} m_{A}(t) + J_{{B}{B}} m_{B}(t) + J_{{B}E} m_E(t))) \\
% m_E(t+1) = \tanh(\frac{1}{3}( J_{E{A}} m_{A}(t) + J_{E{B}} m_{B}(t) + J_{EE} m_E(t)))
% \label{eq:agent-sm}
% \end{aligned}
% \end{equation}
 \begin{equation}
 {\bf m}(t+1) =  \tanh({\bf J m}(t))\\
 \label{eq:agent-sm}
 \end{equation}
where $\bf m$ and $\bf J$ describe in vector and matrix notation the mean fields and couplings of the three regions $A$, $B$ and $E$.
We assume that the environment is defined by two parameters defining the agent environment couplings $J_{AE}=J_{BE}=J_{EA}=J_{EB}=J_c$ and environmental self-couplings $J_{EE}=1$. Values of $J_{{A}{A}},J_{{A}{B}},J_{{B}{A}},J_{{B}{B}}$ will be tuned maximize integration. We also assume $r_S=r_M=r_E=1/3$.

In particular, the system will be tuned to maximize the integrated information of the agent $AB$, $\varphi_{AB}$ while facing 5 different environments defined by values of $J_c$ from the set $\{0.8,0.9,1.0,1.1,1.2\}$.
We calculate $\varphi$ for different parameters as in previous cases, testing the possible candidates for the MIP (in the case of $\varphi_{AB}$, the MIP candidates are isolating one node either from $A$ or $B$) and the one minimizing integrated information is chosen.

In order to find the parameter values that maximize $\varphi_{AB}$ for the set of environments, we first run a microbial genetic algorithm \cite{harvey_microbial_2009} and then (using the parameters of the agent with larger fit) a Nelder-Mead algorithm \cite{nelder_simplex_1965} to adjust the results.
For both algorithms, the fitness function is defined as the value of $\varphi_{AB}(\tau)$, with some exceptions.
For reducing the computational cost, the value of $\tau$ will be $10^4$ for the genetic algorithm and $10^5$ for the Nelder-Mead algorithm. In order to avoid the case where $A$ and $B$ are independent integrated units, fitness will be set to zero in the case that $\varphi_A$ or $\varphi_B$ are larger than $\varphi_{AB}$. As well, fitness is set to zero in the case where $\varphi_{AB}$ does not converge to a stationary value.

After running the genetic and Nelder-Mead algorithms, we obtain an agent with parameters $J_{AA}= 0.09973671, J_{AB}=-0.85774749$, $J_{BA}=-0.8995672$ and $J_{BB}= 0.14326043$. This agent presents negative weights connecting $A$ and $B$ and positive self-coupling values. Thus, each region will inhibit the behaviour of the other while reinforcing itself, therefore regulating its activity to maintain high integrated information for the presented environments.

After tuning the parameters of the system, we evaluate its behaviour for different environments.
For the values of $J_c$ used during training, we find that the mean values of regions $A$ and $B$, $m_A$ and $m_B$ display a similar transition than the previous examples (Fig~\ref{fig:agent-sensorimotor}.B shows the case of $J_c=1$, although other cases are similar).
Moreover, we can observe that there is a divergence of the values of $\varphi_{AB}$ for a range of values of $J_c$ (Fig~\ref{fig:agent-sensorimotor}.C). For larger values of $J_c$ the transition disappears and the values of $\varphi_{AB}$ do not diverge.

The example presented here displays an important qualitative change in comparison with the previous one.
The value of $\varphi_{AB}$ diverges but not only for a specific environment due to fine tuning of its self-couplings as in the previous case. Instead, the divergence is maintained for an approximate range of $J_c$ of $[-1.21,1.21]$. 
Moreover, this divergence is also maintained if we modify the value of $J_{EE}$, displaying a surface in which the value of $\varphi(\tau)$ diverges (Fig~\ref{fig:agent-sensorimotor}.D).
This means that the points of divergence from previous examples are transformed here into a critical surface that maintains integration of the system for a wide range of environmental parameters. That is, the agent is able to self-regulate to some extent to maintain its integration, and thus its viability as an agent.

\section{Discussion}

We have proposed a simplified measure of IIT measure $\varphi$ which, together with mean field approximations in a kinetic Ising model, allows us to capture for the first time integrated information in very large systems, up to the thermodynamic limit. 
Using this method we are able to compute $\varphi$ for infinite size mean field kinetic Ising models with quasi-homogeneous infinite-range connectivity.

Our models, although highly idealized, allow us to speculate about some of the properties of integrated neural organization. First, we observe that, despite the infinite size of the models, the amount of integrated information is bounded for most of its parameter space. Only near critical points does the level of total integrated information diverge, suggesting that integrated entities need to organize themselves close to critical points in their parameter space to maintain their level of integration as their size grows. 
This suggests that it may be of greater interest to describe brain organization in terms of diverging tendencies of IIT in different modules rather than in therms of the specific values of $\varphi$ in finite systems.

Furthermore, we have shown how integrated information can be used to define the boundaries between a system and its environment by comparing the divergent tendencies of their joint and individual integration.
For doing so, some of the assumptions of current formulations of IIT had to be modified. Our tests show that integrated information cannot, in principle, be measured in a brain independently of its environment (bodily and extra-bodily), nor by assuming that the environment is an independent source of noise. Moreover, our results show that near critical points in some cases both the system and system-environment integrated information diverges. Nevertheless, we have shown how to characterize the dominant dynamical unit by comparing the difference in the diverging tendencies between the two configurations.

Our results connect the emergence of boundaries of integration with phenomena related to criticality. Systems near critical points are maximally sensitive to changes in some directions of their parameter space (generally measured as the susceptibility of the system to changes in this parametrical direction). Here, we capture integrated information measures by applying different partitions to the system which are interpreted as changes in particular directions of the parameter space. Thus, the level of integrated information corresponds to the susceptibility of the system for the minimum information partition, i.e., the partition with the less significant effect on the system's causal powers. In the framework of IIT, systems highly sensitive to their minimum information partition are interpreted as maximally irreducible units.

This could allow further simplifications in order to measure integrated information in complex models or even empirical setups. By testing the behaviour of a system when perturbations in its components are introduced (i.e., noise injected in partitioned connections), the integrated information of a mechanism can be described as the minimal susceptibility the set of perturbations from different partitions.
The connection between information integration and critical susceptibility allows us to speculate about the link between integration and properties that have been postulated as pervasive of living beings such as self-organized criticality \cite{bak_self-organized_1988}. 

By interpreting integrated information in terms of susceptibilities in the parametrical direction of partitions of the system, we can think of integration as the sensitivity of a system to the decoupling of the modules composing it. 
In our last example, we show how internal regulation results in the capacity for maintaining this susceptibility for a range of different situations. 
We hypothesize that this can be achieved by similar dynamics as those of systems showing self-organized criticality, which are attracted to critical points of maximum susceptibility. This could be achieved in systems capable of self-organizing near points where they maintain maximal sensitivity to the integrity of their internal organization while they interact with changing environments (e.g., maintaining internal invariances near critical surfaces \cite{aguilera_adaptation_2018}).

\section{Conclusion}

The core ideas that IIT intends to capture apply to a variety of poorly understood questions in biological and cognitive systems. By introducing some modifications to take into account different temporal spans  and influences from the environment, and studying the behaviour of integration measures in the thermodynamic limit, we have shown the existence of critical points that maximise a system's integration, for instance, an organism or a cognitive agent. The fact that our case studies remain general and abstract (we do not specify any detail about the neural,  sensorimotor, and environmental processes involved) suggests that robust individuation and susceptibility towards loss of integration are inherent consequences of maximising a tendency towards integration, and so they are likely to be observable trends in all systems that are able to do so.

A limiting assumption in our approach is the homogeneity of the elements within a each region. Biological systems cannot be assumed to present such a degree of homogeneity and the variability in their components and interactions has to be accounted for. Our framework, however, can take into account higher levels of heterogeneity by introducing a larger number of regions. In the case of three regions we observe that tuning the parameters of the system results in the extensions of critical points of diverging integration into regions of the parameter space. We expect (but have not yet verified) that increasing the number of interacting regions will still result in critical regions of divergent integration. In brain network models, it has been found that structural heterogeneity can generate extended critical-like regions \cite{moretti_griffiths_2013}, thus we may also expect this phenomenon to be reinforced in the presence of higher heterogeneity in our models. Our results are also limited to models with stationary solutions where we can evaluate the stable solution when the temporal span tends to infinity. This is not a limitation of the method, though. The results of more realistic systems presenting cyclic or chaotic dynamics could be harder to interpret, although they are in principle tractable within the framework presented here and could be explored in further work.

The models presented here allow a shift of focus toward the integrative tendencies of systems as they grow or evolve. This opens up the applicability of IIT to a range of questions about changes over developmental and evolutionary time. Even in the simple cases we have considered, the existence of critical points that maximise integration may be important for understanding apparent jumps in complexity, including the transitions at the origin of life \cite{walker_algorithmic_2013} or cognitive developmental transitions \cite{molenaar_commentary_2004}.

Focusing on the divergent tendencies of integration measures, we are able to capture the asymmetry of agent-environment interactions. Thinking interactions with the environment in this terms is fruitful for grounding notions such as the individuality or the autonomy of a system. Often, these concepts have been formalized in terms of self-determination and independence from an environment \cite{bertschinger_autonomy:_2008, krakauer_information_2014}. By contrast, our examples show how both integration of a system and integration between system an environment can diverge together, while the level of individuality of the system can be quantified by the relative divergence speed of both terms. This is a robust finding obtained under the minimal assumptions  and thus, we suggest, a general trend in large complex systems. The key data of interest as systems scale up are not so much the absolute values of integrated information, but the relative divergent tendencies of system integration and system-environment integration.

In addition, by exploring different kinds of agent-environment configurations, we observe that agents assumed to maximise integration are likely to do so robustly for a range of environmental situations due to the existence of critical surfaces. The existence of these surfaces that guarantee maximal integration is coherent with postulates at the theoretical foundations of adaptive systems research, such as the existence of `regions of viability’ that guarantee the integrity of an agent \cite{ashby_design_1960, barandiaran_norm-establishing_2014}. While such conditions of viability have often been imposed by the designer or assumed to be given by evolutionary or material constraints, our approach allows to think of them as critical regions emerging at the level of the integrative forces of the system. This illustrates how viability regions could scale up from material or pre-given constraints to regions defined by increasing complexity of the integrated activity of a system.

\begin{acknowledgments}
M.A. was supported by the \mbox{UPV/EHU} post-doctoral training program \mbox{ESPDOC17/17} and project \mbox{TIN2016-80347-R} funded by the Spanish Ministry of Economy and Competitiveness.
\end{acknowledgments}

\appendix

\section{IIT 3.0}
\label{app:IIT}
In the last version  of integrated information theory \cite{oizumi_phenomenology_2014}, integrated information of a subset of elements of a system is computed as follows. 
For a system of elements $S$ in state $s$, we describe the input-output relationship of the system elements through  its  corresponding  transition  probability  function  $p$, describing  the probabilities  of the transitions  from  one  state  to  another  for  all  possible  system states.
IIT requires that $p$ satisfies the Markov property (i.e., the state at time $t$ only depends on the state at time $t-1$), and that the  current states  of  elements  are  independent,  conditional  on  the  past  state  of  the  system. This conditions are satisfied by the asymmetric kinetic Ising model used here. 

For any two subsets of $S$,  called the  mechanism $\mathcal{M}$  and the  purview  $\mathcal{P}$,  we  can define the cause and effect repertoires of $\mathcal{P}$ over $\mathcal{M}$, that is, how $\mathcal{M}$ in its current state $\{s_i(t)\}_{i\in\mathcal{M}}$, constrains the potential past or future states of $\{s_i(t-1)\}_{i\in\mathcal{P}}$ or $\{s_i(t+1)\}_{i\in\mathcal{P}}$. Cause and effect repertoires of the system are described by the probability distributions $p(\mathcal{P}_{t-1} |\mathcal{M}_{t} ) = p(\{s_i(t-1)\}_{i\in\mathcal{P}} | \{s_i(t)\}_{i\in\mathcal{M}})$ and $p(\mathcal{P}_{t+1} |\mathcal{M}_{t} ) = p(\{s_i(t+1)\}_{i\in\mathcal{P}} | \{s_i(t)\}_{i\in\mathcal{M}})$.

The integrated cause-effect information of $\mathcal{M}$  is then defined as the distance between the cause-effect repertoires of the mechanism, and the cause-effect repertoires of their minimum information partition (MIP) over the purview that is maximally irreducible,
\begin{equation}
\begin{aligned}
\varphi_{cause}= \underset{\mathcal{P}}{max}\left (\underset{cut}{min}\left (D(p(\mathcal{P}_{t-1} | \mathcal{M}_{t}),p^{cut}(\mathcal{P}_{t-1} | \mathcal{M}_{t}))\right )\right ) \\
\varphi_{effect}= \underset{\mathcal{P}}{max}\left (\underset{cut}{min}\left (D(p(\mathcal{P}_{t+1} | \mathcal{M}_{t}),p^{cut}(\mathcal{P}_{t+1} | \mathcal{M}_{t}))\right )\right ) 
\end{aligned}
\end{equation}
where $cut$ is a partition of the mechanism into two halves, and $p^{cut}$ the cause or effect probability distribution under the partition,
\begin{equation}
\begin{aligned}
cut = \left \{ \mathcal{M}_1,\mathcal{P}_1,\mathcal{M}_2,\mathcal{P}_2 \right \} \\
p^{cut}(\mathcal{P} |\mathcal{M}) = p(\mathcal{P}_1 | \mathcal{M}_1)\otimes  p(\mathcal{P}_2 | \mathcal{M}_2)
\end{aligned}
\end{equation}

The  integrated  information  of  the  mechanism  $\mathcal{M}$  is  the  minimum  of  its  corresponding  integrated  cause  and  effect information, 
\begin{equation}
\varphi = min(\varphi_{cause},\varphi_{effect})
\end{equation}

The  integrated  information  of  the  entire  system  is  then  defined  as  the  distance  between  the  cause-effect structure  of  the  system,  and  cause-effect  structure  defined  by  its  minimum  information  partition,  eliminating constraints from one part of the system to the rest:
\begin{equation}
    \Phi = \underset{cut}{min}~D(C,C^{cut})
\end{equation}

For both the integrated information of a mechanism ($\varphi$) and the integrated information of a system ($\Phi$), distance $D$ is computed as the Wasserstein or earth mover’s distance. Finally, if $S$ is a subset of elements of a larger system, all elements outside of $S$ are considered as part of the environment and are conditioned on their current state throughout the causal analysis. 
Further details of the steps described here can be found in reference [6]
%\cite{oizumi_phenomenology_2014}.

\section{Simplified integrated information $\varphi$}
\label{app:phi}

\setcounter{figure}{0}
\renewcommand{\thefigure}{\ref{app:phi}\arabic{figure}}%
\renewcommand{\thesubsection}{\ref{app:phi}\arabic{subsection}}
Measures in this paper are inspired by the IIT framework, although we apply some modifications and simplifications.

\subsection{Temporal range}
\label{app:phi-time}
First, as we mentioned in the paper, we only compute the value of $\varphi$ for the effects of the current system in a posterior state $t+\tau$, while IIT computes the minimum of $\varphi_{cause}$ and $\varphi_{effect}$ at $t-1$ and $t+1$.
However, IIT can also deal with temporal scales. As IIT operates with the transition probability matrix of a system, one could compute this matrix from time $t$ to time $t+\tau$ and apply the operations for computing $\varphi$ over it. This implies that the noise injected by partitions in the connections that are cut down is only injected at time $t$, and the system behaves normally for the following steps. In our case, we inject independent noise at every update from time $t$ to $t+\tau$.

We can test the difference between the two approaches in a homogeneous kinetic Ising model with $H_i=0$ and $J_{ij}=J$. As we showed in the paper, applying a continuous noise injection in partitions makes the value of $\varphi$ diverge around the critical point $J=1$ as $\tau$ grows (Figure \ref{fig:iit-external-activity}.A).
Conversely, in we only apply an initial noise injection at partitioned connections, we see that the measured $\varphi$ operates in a different way (Figure \ref{fig:iit-external-activity}.B). In this case, as $\tau$ increases, the value of $\varphi$ decreases as the system regains stability in its original position. Moreover, for small values of $\tau$ the values of $J$ with larger $\varphi$ are above the critical point.
However, we observe that, the closer we are to the critical point, the slower $\varphi$ decreases. This is due to a phenomena called `critical slowing down', a phenomena characteristic of critical dynamics in which the response time of a system near criticality tends to infinity. Curiously, if we compute the cumulative sum of the values of $\varphi$ from 1 to $\tau$, i.e. $\varphi_{cum} = \sum\limits_{\tau'=1}^{\tau} \varphi(\tau')$ (Figure \ref{fig:iit-external-activity}.C), we observe that the result is identical to the case of continuous noise injection at partitions.

\begin{figure*}
\includegraphics[width=16.0cm]{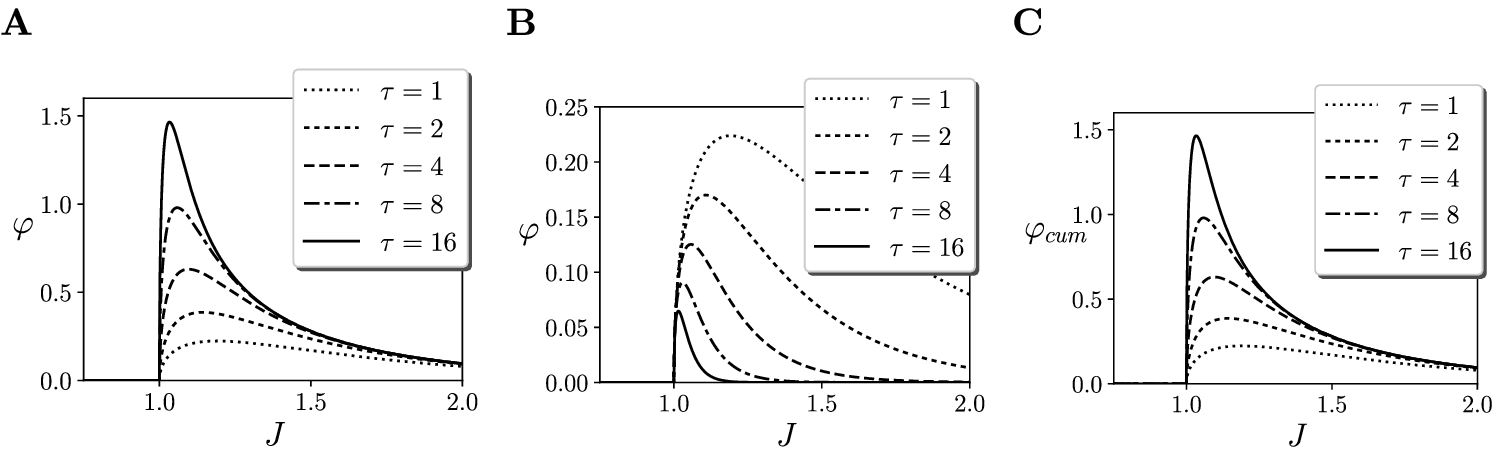}
\caption{Temporal ranges of integration. \textbf{(A)} Values of $\varphi(\tau)$ using continuous injection of noise for different values of $J$. \textbf{(B)} Values of $\varphi(\tau)$ using an initial injection of noise for different values of $J$. \textbf{(C)} Values of $\varphi_{cum} = \sum\limits_{\tau'=1}^{\tau} \varphi(\tau')$  using an initial injection of noise for different values of $J$.} 
\label{fig:iit-time}
\end{figure*}

\subsection{Purview}

In IIT, integrated information of a mechanism $\varphi^{MIP}_\mathcal{M}$ is evaluated not only for a particular mechanism $\mathcal{M}$, but also for a purview $\mathcal{P}$. If the mechanism defines which units of $\{s_i(t)\}_{i \in \mathcal{M}}$ we take into account, the purview defines which units of the future state $\{s_i(t+\tau)\}_{i \in \mathcal{P}}$ we take into account. Given these subset of present and future states, partitions are computed over the join space of $\{s_i(t)\}_{i \in \mathcal{M}}$ and $\{s_i(t+\tau)\}_{i \in \mathcal{P}}$, and the purview $\mathcal{P}$ with maximum integrated information for its MIP is selected.
Here for simplicity, we apply the partition over $\{s_i(t)\}_{i \in \mathcal{M}}$ and $\{s_i(t+\tau)\}_{i \in \mathcal{M}}$, making the mechanism and purview coincide, and the distance for computing integrated information is measured for the distance of all elements of the system, not only the elements contained in the purview.

Allowing more choices of purview could make a big difference in certain systems, although in the quasi-homogeneous systems tested in the paper the differences are small.

\subsection{Elements outside of a mechanism}
\label{app:phi-environment}
More importantly, there are significant differences from the IIT framework in the way we treat the elements that are outside of the evaluated mechanism $\mathcal{M}$.
In IIT, elements outside the mechanism are assumed to be unconstrained (i.e., as random as possible).
We decided to modify this assumption because it can have dramatic effects when measuring the behaviour of large systems. Specifically, assuming unconstrained elements outside the mechanism create an artifact that provokes a shift in the critical point of the system (this will be detailed in future work).

Let's exemplify an example using an homogeneous Ising model with local fields $H_i=0$ and couplings $J_{ij}=J$.
As we shown,  compute the value of $\varphi$ for the whole system using continuous noise injection at partitioned connection yields a divergence around the critical point at $J=1$. Now, we will show what is the behaviour of its internal mechanisms assuming different behaviours of the units outside of the mechanism.

\begin{figure*}
\includegraphics[width=16.0cm]{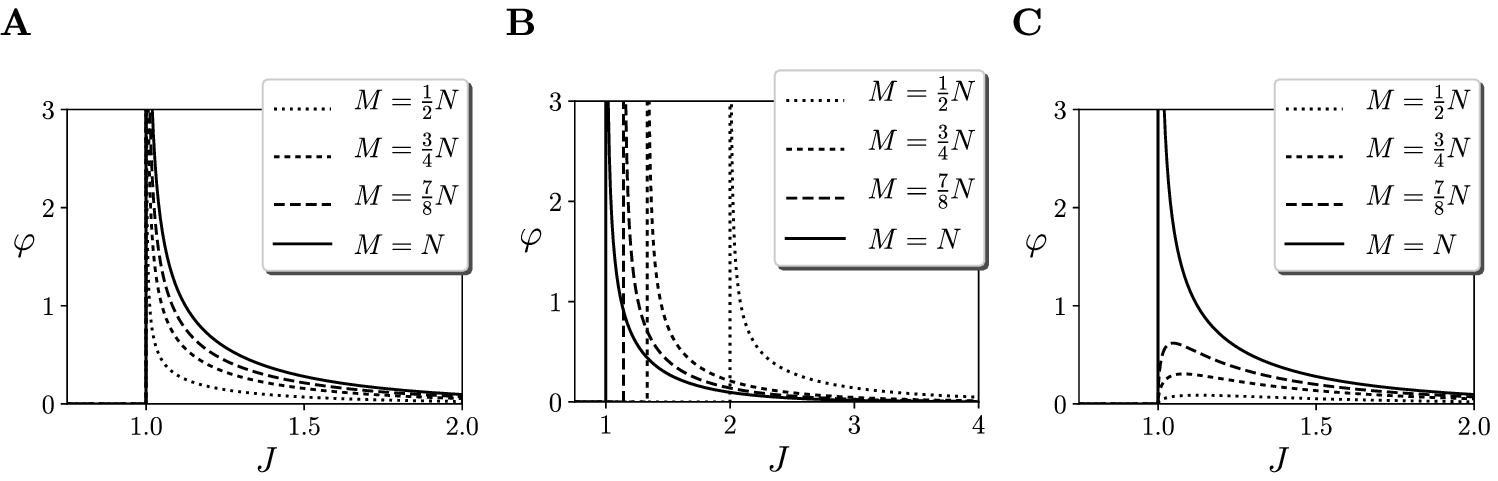}
\caption{Effects of the environment in integrated information. Values of $\varphi_\mathcal{M}(\tau\to\infty)$ of a mechanism $\mathcal{M}$ of size $M$ for different values of $J$, assuming that elements outside of the mechanism operate (A) normally, (B) as independent sources of noise, and (C) as static input fields.} 
\label{fig:iit-external-activity}
\end{figure*}

First, we compute values of mechanism covering a fraction of the system $M/N$ (since the system is homogeneous, any fraction we choose has the same behaviour) assuming that the elements outside of the mechanism $\mathcal{M}$ keep operating normally (Figure~\ref{fig:iit-external-activity}.A). In this case, we observe that the divergence of $\varphi_\mathcal{M}$ is maintained, although the value of  $\varphi_\mathcal{M}$ decreases with the mechanism size.

In contrast, if we accept IIT assumption and take the elements of the mechanism as independent sources of noise, the behaviour of $\varphi_\mathcal{M}$  changes radically. In this case, the divergence is maintained but takes place at a different value of the parameter $J$ (Figure~\ref{fig:iit-external-activity}.B). This happens because independent sources of noise have a zero mean field value, and thus the phase transition of the system takes place at larger values of $J$ that compensate the units that now are contributing with a zero mean field.
Thus, we think that considering the elements outside of the mechanism as independent sources of noise can be misleading about the operation of mechanisms that are embedded in large systems.

A less loaded assumption could be maintaining the state of the units outside of the mechanism with the static values that they had at time $t$, that is, maintaining their mean field constant. We can see at Figure~\ref{fig:iit-external-activity}.C that this behaviour is also not satisfactory, since for mechanism sizes smaller than $N$ the value of $\varphi_\mathcal{M}$ decreases very rapidly, and it is exactly zero at the critical point. We can understand this thinking that the effect of constant fields is equal to adding a value of $H_i$ equal to the input from frozen units, therefore breaking the symmetry of the system and precluding a phase transition.

\subsection{Mean field approximation of partitioned systems}

We simplify the calculation of the probabilities $p(\{s_i(t+\tau)\}_{i \in \mathcal{M}}|\{s_i(t)\}_{i \in \mathcal{M}}) $ and $p^{cut}\{s_i(t+\tau)\}_{i \in \mathcal{M}}|\{s_i(t)\}_{i \in \mathcal{M}})$ by using a mean field approximation described by Equations~\ref{eq:mean-field}~and~\ref{eq:update-mean-field}. 

In the case of partitioned systems for computing integrated information, cutting connections injects uniform noise on the input node. 
In the mean field approximation, this would be equivalent to inject a zero mean field signal, which is equivalent to setting to zero the affected connection weights when computing $h_i(t)$.

\subsection{Integrated conceptual information}

Finally, once $\varphi$ is computed, IIT proposes a second level of calculations for computing integrated conceptual information $\Phi$ where new bidirectional partitions are applied to the system. 
In our case, given the homogeneity of the system, we do not compute conceptual information since all the mechanisms composing each set have similar behaviour. Thus, for simplicity  we do not apply a second level of partitions.

\bibliography{references}% Produces the bibliography via BibTeX.

\end{document}